\documentclass[conference]{IEEEtran}

\usepackage[T1]{fontenc}

\usepackage{ifpdf}

\usepackage{cite}

\ifCLASSINFOpdf
  \usepackage[pdftex]{graphicx}
   \graphicspath{{../pdf/}{../jpeg/}}
   \DeclareGraphicsExtensions{.pdf,.jpeg,.png}
\else
\fi
\usepackage{amsmath}
\usepackage{algorithmic}

\usepackage{array}
\newcolumntype{C}[1]{>{\centering\let\newline\\\arraybackslash\hspace{0pt}}m{#1}}

\usepackage{color}
\definecolor{dblue}{rgb}{0,0,0.8}

\usepackage{xcolor}

\usepackage{subfig}
\usepackage{graphicx}
\usepackage{comment}

\usepackage{stfloats}
{}
{}
{}

\usepackage{booktabs, multicol, multirow}
\usepackage{tabularx}
\usepackage{lipsum}

\begin{document}


\title{Proactive Scheduling of Hydrogen Systems for Resilience Enhancement of Distribution Networks}





%
%
%

\author{\IEEEauthorblockN{Hamed Haggi}
\IEEEauthorblockA{Department of Electrical and\\Computer Engineering\\
University of Central Florida\\
Orlando, Florida, USA\\
hamed.haggi@knights.ucf.edu}
\and
\IEEEauthorblockN{Wei Sun}
\IEEEauthorblockA{Department of Electrical and\\Computer Engineering\\
University of Central Florida\\
Orlando, Florida, USA\\
sun@ucf.edu}
\and
\IEEEauthorblockN{James M. Fenton}
\IEEEauthorblockA{Florida Solar Energy \\ Center \\
University of Central Florida\\
Cocoa, Florida, USA\\
JFenton@fsec.ucf.edu}
\and
\IEEEauthorblockN{Paul Brooker}
\IEEEauthorblockA{Orlando Utilities Commission\\
Orlando, Florida, USA\\
PBrooker@ouc.com}
}

\maketitle


\begin{abstract}
Recent advances in smart grid technologies bring opportunities to better control the modern and complex power grids with renewable integration. The operation of power systems, especially distribution network (DN), is facing with preeminent challenges from cyber-physical-human (CPH) threats and natural disasters. In order to provide better response against threats and improve the resilience of power grid, proactive plans and operational schemes are required by system operators to minimize the damages caused by CPH threats. To that end, this paper proposes a proactive plan for DN operation by using hydrogen (H2) systems to enhance the resilience through cost-effective long-term energy storage. Unlike batteries, H2 energy can be stored in the storage tanks days before the extreme event, and transformed into power by fuel cell units in the post-event time to reduce load curtailment caused by CPH threats. The proposed framework is validated by testing on 33-node test feeder. Simulation results demonstrate that H2 systems can improve the resilience of DN during $N-m$ outages lasting for more than 10 hours. \par
\end{abstract}
\vspace{0.25cm}

%
\IEEEpeerreviewmaketitle

\section{Introduction}
\IEEEPARstart{R}{ecently}, due to the large deployment of distributed energy resources (DERs) and other smart grid technologies, power systems are becoming more complex and vulnerable to extreme events. These extreme events can significantly affect the operation of power systems and results in outages and even cascading failures \cite{khazeiynasab2020resilience}. For instance, Hurricane Sandy (October 22-November 2, 2012) affected 7.5 million electricity consumers (mainly on the distribution side) with \$65 billion in damage\cite{gholami2017proactive}. Another example is the extreme cold temperature in the state of Texas (February 2021) affected the social and economic welfare of 4.5 million consumers \cite{DOE}. To learn from the past events, utilities must pay extra attention to the resilience enhancement plans before the events. The promising strategies that the distribution system operator (DSO) can take to improve the resilience of distribution network (DN) include, 1) utilizing DERs to provide back up energy for serving local or critical loads; 2) damage preventive actions or system hardening; and 3) automation technologies to enhance the responsiveness to outages, etc. \cite{haggi2019review}. \par

Recent research works have been focused on various aspects of resilience enhancement, such as proactive energy management, survivability analysis, and restoration actions. Since this paper is only focused on proactive scheduling as well as survivability analysis, restoration papers are not reviewed. However, interested readers are encouraged to refer to our previous work of reviewing recent advances in smart grid restoration \cite{haggi2019review}. Authors of \cite{nguyen2020preparatory} proposed a preparatory operation scheme for resilience enhancement of critical loads in automated DN. The framework was modeled as a stochastic optimization program considering pre- and post-event actions. A proactive operation scheme was proposed in \cite{hussain2018proactive} considering battery energy storage while addressing the uncertainty of load and renewable generation. In \cite{gholami2017proactive}, a two-stage adaptive robust framework for resilience improvement with microgrids was proposed. A proactive scheduling considering the benefits of microgrid for resilience enhancement against windstorms was presented in \cite{amirioun2017resilience}. The proposed method considered conservation voltage reduction, optimal parameter settings of droop controlled units and reconfiguration.\par  

As mentioned above, proactive management of DERs can enhance the resilience of power systems. Recently, hydrogen (H2) energy has gained a lot of attention and has demonstrated a great potential for large-scale and long-duration energy storage deployment in the near future because of its economic, environmental, and technical merits. For more information regarding the H2 energy safety, interested readers are encouraged to see \cite{barilo2017overview}. Unlike batteries which can only store/discharge energy for 4-8 hours with its maximum power and energy rating, H2 storage can store/discharge with maximum discharging capability for longer period of time (e.g. days and months). Research works in the literature are mainly focused on the normal operation of H2 systems and market mechanisms for H2 trading. For instance, a supervisory scheduling of H2 refueling stations considering the operating reserve signals was proposed in \cite{khani2019supervisory}. An optimal scheduling of H2 refuelling stations with the aim of fuel supply and capacity-based demand response for electrolyzers was analyzed in \cite{el2018hydrogen}. A decentralized local market for electricity and H2 trading based on game theory was proposed in \cite{xiao2017local} by considering the privacy of agents. An optimal energy management framework was proposed in \cite{rahmanzadeh2018optimal} considering H2 energy as a fuel for proton exchange membrane fuel cell (FC) units. In this paper, H2 energy can be transformed to power with FC units and assist the grid in supplying the electric and thermal load. \par 

Proactive scheduling as well as survivability analysis considering H2 systems to improve the DN resilience have not been fully investigated yet. Additionally, since the DSO is not the owner of all assets in DN, proper models are required for sending signals to H2 system owners in order to get prepared for emergency operations. To that end, the major contribution of this work is the proactive scheduling (pre-event analysis) and survivability analysis (post-event analysis) of H2 systems to improve the DN resilience. Prior to disruptions (e.g. days before), DSO sends signals to H2 system owners and asks them to completely fill their storage tank. This results in maximizing the survivability by using the stored H2 for stationary FC consumption and consequently the resilience improvement. Additionally, deploying long-term H2 storage results in lower cycling of DGs which increases the operational life of DGs and improves the resilience of these assets in the long term.\par

The rest of the paper is organized as follows. The problem formulation is introduced in Section \ref{Formulation}, simulation results are presented in Section \ref{results}, and Section \ref{conclusion} concludes the paper.\par

\section{Problem Formulation}\label{Formulation}
The mathematical formulation of proactive scheduling of H2 systems in active distribution networks is presented in this section. The proposed model is formulated as a resilience constrained program (RCP) to address both proactive scheduling and survivability analysis. Given a DN, $(N,l)$, where $N$ and $l$ are the set of nodes and lines, the root node is shown as 0, and $T$ represents the set of time steps indexed by $t$.   

\subsection{Objective Function}
The objective function of the model is to minimize total operation cost of DN throughout the entire period of $T$ which can address both normal and emergency operation of DN. 

\begin{equation}\label{OF}
\begin{split}
\text{min.} & \sum_{t=1}^{T}\Bigg\{\rho_t P^{ST}_t + \sum_{i=1}^{N_G} C^{DG}_{i,t} + \sum_{i=1}^{N_G} C^{SU}_{i,t} + \sum_{i=1}^{N_G} C^{SD}_{i,t} \\ & + \sum_{i=1}^{N_{PV}} C^{PV}_{i,t} + \sum_{i=1}^{N} C^{Shd}_{i,t}  \Bigg\}
\end{split}
\end{equation}

where the first term refers to the cost of purchasing power from upper grid with the price of $\rho_t$. The second, third, and forth terms refer to the operational cost of utility operated power plants, start up and shutdown costs, respectively. The last two cost terms refer to the cost of utility operated PV units and load curtailment, respectively. It should be noted that the costs associated with H2 systems are not taken into account and DSO only schedules its system demand. In this paper, both DSO and H2 system owners perform the cost benefit analysis separately and exchange energy with power purchase agreement price \cite{shekari2016techno}. \par

\subsection{Technical Constraints and Models}
The aforementioned objective function is subjected to the following operational constraints for DGs, PVs, and H2 systems including electrolyzer, storage tank and stationary FC units.\par

\subsubsection{Operational Constraints for DG Units}
Let $x^{DG}_{i,t}$, $P^{DG}_{i,t}$, $Q^{DG}_{i,t}$, $b^{DG}$, $k^{DG}$, and $C^{DG}_{i,t}$ denote to the status, active power, reactive power, fixed operation and maintenance cost, ramping cost, and operational cost of DGs, respectively. Constraint (\ref{DGcost}) consists of two terms, fixed operation cost and ramping cost of DGs. Constraints (\ref{PDGlimit})-(\ref{DGSOCP}) show the active and reactive power production limits. Startup cost of DGs, $C^{SU}_{i,t}$, as well as their shutdown cost, $C^{SD}_{i,t}$, are expressed in (\ref{DGSU1})-(\ref{DGSD2})\cite{gholami2016microgrid}. Moreover, the ramp rate limits of DG are shown in (\ref{DGramp}), which $R^D_i$ and $R^U_i$ show ramp-down and ramp-up limits, respectively \cite{haggi2017security}.
\begin{equation}\label{DGcost}
C^{DG}_{i,t} = x^{DG}_{i,t}.\; b^{DG}+ k^{DG}.\;P^{DG}_{i,t}
\end{equation}
\begin{equation}\label{PDGlimit}
P^{DG,min}_{i,t}.\;x^{DG}_{i,t} \le P^{DG}_{i,t} \le P^{DG,max}_{i,t}.\;x^{DG}_{i,t}
\end{equation}
\begin{equation}\label{QDGlimit}
Q^{DG,min}_{i,t}.\;x^{DG}_{i,t} \le Q^{DG}_{i,t} \le Q^{DG,max}_{i,t}.\;x^{DG}_{i,t}
\end{equation}
\begin{equation}\label{DGSOCP}
(P^{DG}_{i,t})^2 + (Q^{DG}_{i,t})^2 \le (S^{DG})^2
\end{equation}
\begin{equation}\label{DGSU1}
C^{SU}_{i,t} \ge (x^{DG}_{i,t}-x^{DG}_{i,t-1}).\;\rho^{SU}
\end{equation}
\begin{equation}\label{DGSU2}
C^{SU}_{i,t} \ge 0
\end{equation}
\begin{equation}\label{DGSD1}
C^{SD}_{i,t} \ge (x^{DG}_{i,t-1}-x^{DG}_{i,t}).\;\rho^{SD}
\end{equation}
\begin{equation}\label{DGSD2}
C^{SD}_{i,t} \ge 0
\end{equation}
\begin{equation}\label{DGramp}
-R^D_i \le P^{DG}_{i,t} - P^{DG}_{i,t-1} \le R^U_i
\end{equation}
It is worth mentioning that other constraints such as minimum uptime and minimum downtime of DGs can be considered based on \cite{haggi2017security} to have more realistic results due to system inflexibility caused by DGs in some hours, which increases the total operation cost. To address this inflexibility, large scale H2 systems can be considered to reduce the total cost. 
\subsubsection{Operational Constraints for PVs and Substation Node}
The operational cost of PV units, $C^{PV}_{i,t}$, is expressed in (\ref{PVcost}). $P^{PV}_{i,t}$ and $Q^{PV}_{i,t}$ are denoted as active and reactive power of PV units, and their limits are expressed in (\ref{PVlimit}) and (\ref{PVreactive}) \cite{rahmani2021cool}. More information regarding the smart PV inverters and their role on resilience enhancement can be referred to \cite{fard2020multitimescale} \cite{fard2020holistic}. Additionally, the active and reactive power purchased from upper grid, $P^{ST}_{i,t}$ and $Q^{ST}_{i,t}$, are constrained by (\ref{uppergrid}). 
\begin{equation}\label{PVcost}
C^{PV}_{i,t} = c^{PV}.\;P^{PV}_{i,t}
\end{equation}
\begin{equation}\label{PVlimit}
0 \le P^{PV}_{i,t} \le P^{PV, max}
\end{equation}
\begin{equation}\label{PVreactive}
(P^{PV}_{i,t})^2 + (Q^{PV}_{i,t})^2 \le (S^{PV})^2
\end{equation}
\begin{equation}\label{uppergrid}
(P^{ST}_{i,t})^2 + (Q^{ST}_{i,t})^2 \le (S^{ST})^2
\end{equation}

\subsubsection{Operational Constraints for H2 System}
The operational constraints for H2 systems including electrolyzers, storage tank, and stationary FC units are expressed in (\ref{ELX})-(\ref{DRMOH}). Let us denote $P^{EL}_{i,t}$,$P^{FC}_{i,t}$, $Q^{EL}_{i,t}$, and $Q^{FC}_{i,t}$ to electrolyzer power, FC power, amount of H2 consumed by electrolyzer, and generated power by FC units, respectively. Constraints (\ref{ELX}) and (\ref{FC}) show the amount of H2 produced or consumed by electrolyzers or FCs, based on efficiency $\eta$, power to H2 $\lambda^{EL}$, H2 to power $\lambda^{FC}$, and converting factors \cite{khani2019supervisory}-\cite{el2018hydrogen}. Constraints (\ref{ELXlimit}) and (\ref{FClimit}) present the limits for H2 production and consumption, based on the capacity limit of electrolyzers and FC units. Additionally, these constraints prevent the simultaneous operation of electrolyzer and FC units by considering a binary variable $\psi^{HS}_{i,t}$. $MOH^{H2}_{i,t}$ denotes to the mass of H2 in storage tank, and dynamic H2 mass constraints are presented in (\ref{MOH}) and (\ref{MOHlimit}). It should be noted that $Q^{dem}_{i,t}$ and $\lambda^{Dsp}$ are H2 demand from fuel cell electric vehicles (FCEVs) and H2 storage dissipation rate. In the case of $N-m$ disruptions, H2 systems can act as long-term energy storage with long-duration times compared to batteries. To that end, constraint (\ref{DRMOH}) express the demand response (DR) signal, in which $\alpha_t$ denotes the percentage of H2 required as a reserve before emergency operation, from DSO regarding the H2 mass in the tank. Prior to any forecasted disruption ($t < t_{event}$), DSO asks H2 system owners to fill their tank completely as a backup generation unit for supplying the load in the post-event time ($t \ge t_{event}$). This will help the DSO to minimize the total cost and total load curtailment during the $N-m$ contingencies. Moreover, constraint (\ref{HSinverter}) expresses the H2 systems inverter for voltage control.  
\begin{equation}\label{ELX}
Q^{EL}_{i,t}= \lambda^{EL}.\; P^{EL}_{i,t} .\;\eta^{EL}
\end{equation}
\begin{equation}\label{FC}
P^{FC}_{i,t}= \lambda^{FC}.\; Q^{FC}_{i,t} .\;\eta^{FC}
\end{equation}
\begin{equation}\label{ELXlimit}
Q^{EL,min}.\;\psi^{HS}_{i,t}\le Q^{EL}_{i,t} \le Q^{EL,max}.\;\psi^{HS}_{i,t}
\end{equation}
\begin{equation}\label{FClimit}
Q^{FC,min}.\;(1-\psi^{HS}_{i,t}) \le Q^{FC}_{i,t} \le Q^{FC,max}.\;(1-\psi^{HS}_{i,t})
\end{equation}
\begin{equation}\label{MOH}
\begin{split}
MOH^{H2}_{i,t}= & MOH^{H2}_{i,t-1}\;+(Q^{EL}_{i,t} - Q^{dem}_{i,t} - Q^{FC}_{i,t} \\ & -\lambda^{Dsp}.\; MOH^{H2}_{i,t})).\;\Delta t
\end{split}
\end{equation}
\begin{equation}\label{MOHlimit}
MOH^{H2, min}\le MOH^{H2}_{i,t} \le MOH^{H2, max}
\end{equation}
\begin{equation}\label{DRMOH}
 \sum_{i=1}^{N_H} MOH^{H2}_{i,t} \ge \alpha_{t}\;.\;\sum_{i=1}^{N_H} MOH^{H2,max}
\end{equation}
\begin{equation}\label{HSinverter}
(P^{EL}_{i,t}-P^{FC}_{i,t})^2 + (Q^{HS}_{i,t})^2 \le (S^{HS})^2
\end{equation}

\subsubsection{Distribution System Constraints}
The AC power flow constraints of DN are shown in (\ref{voltage})-(\ref{shedreactive}), where $V_{i,t}$, $f^p_{i,t}$, and $f^q_{i,t}$ refer to the squared voltage magnitude, active and reactive power flow of lines, respectively. It should be noted that LinDistflow \cite{baran1989network} \cite{taghavirashidizadeh2020genetic} is adopted for modeling the radial distribution system. Constraints (\ref{voltage}) and (\ref{voltagelimit}) express the relationship between voltage drop based on active and reactive power flow in the lines, and voltage limits for every node, respectively. Constraints (\ref{activebalance}) and (\ref{reactivebalance}) show the nodal active and reactive power balance. Constraint (\ref{linelimit}) expresses the line capacity limit for active and reactive power flow. Constraint (\ref{shedcost}) shows the load curtailment cost in the emergency operation mode, which is penalized with the value of loss of load ($VOLL$), which is considered as \$500/MWh in this paper. Finally, constraints (\ref{shedlimit}) and (\ref{shedreactive}) show the limits for load curtailment.
\begin{equation}\label{voltage}
V_{i,t} = V_{j,t} - 2(R_{ji}\;.\;f^p_{i,t} -X_{ji}\;.\;f^q_{i,t})
\end{equation}
\begin{equation}\label{voltagelimit}
(V^{min})^2 \le V_{i,t} \le (V^{max})^2
\end{equation}
\begin{equation}\label{activebalance}
f^p_{i,t} = P^{Load}_{i,t} + \sum_{j\rightarrow i} f^p_{j,t} + P^{EL}_{i,t} - P^{FC}_{i,t} - P^{PV}_{i,t} - P^{DG}_{i,t} - P^{Shd}_{i,t}    
\end{equation}
\begin{equation}\label{reactivebalance}
f^q_{i,t} = Q^{Load}_{i,t} + \sum_{j\rightarrow i} f^q_{j,t} + Q^{EL}_{i,t} - Q^{FC}_{i,t} - Q^{PV}_{i,t} - Q^{DG}_{i,t} - Q^{Shd}_{i,t}     
\end{equation}
\begin{equation}\label{linelimit}
(f^p_{i,t})^2 + (f^q_{i,t})^2 \le (S^{line})^2
\end{equation}
\begin{equation}\label{shedcost}
C^{Shd}_{i,t} = VOLL .\;P^{Shd}_{i,t}
\end{equation}
\begin{equation}\label{shedlimit}
0 \le P^{Shd}_{i,t} \le P^{Load}_{i,t}
\end{equation}
\begin{equation}\label{shedreactive}
Q^{Shd}_{i,t} = P^{Shd}_{i,t}\;.\;\frac{Q^{Load}_{i,t}}{P^{Load}_{i,t}}
\end{equation}
\begin{figure}
\centering
\footnotesize
\captionsetup{justification=raggedright,singlelinecheck=false,font={footnotesize}}
	\includegraphics[width=3.5in,height=1.65in]{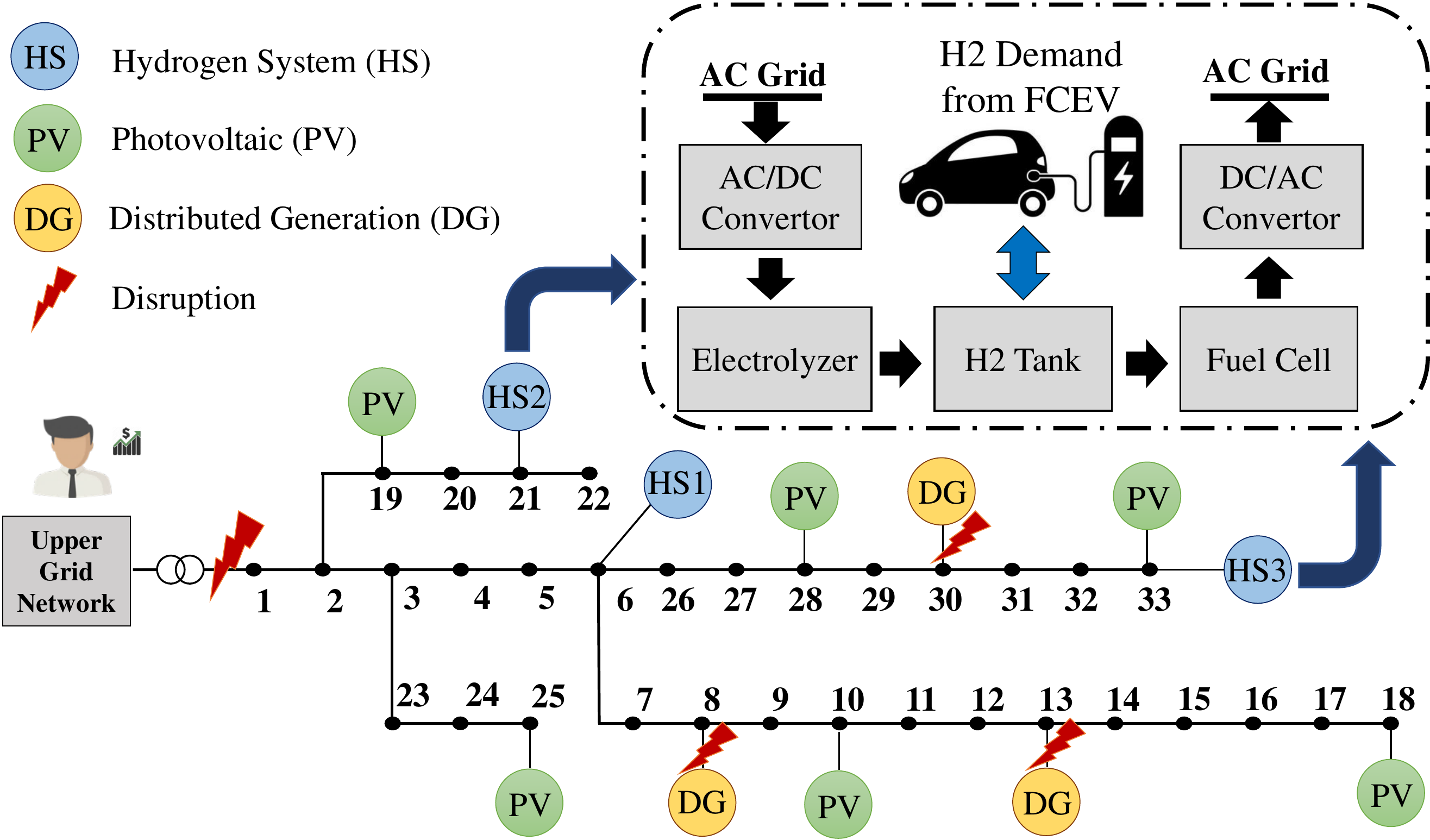}
	\caption{Radial 33-node distribution test system with DERs.}
    \label{system}
\end{figure}
\begin{figure}
\centering
\footnotesize
\captionsetup{justification=raggedright,singlelinecheck=false,font={footnotesize}}
	\includegraphics[width=3.5in]{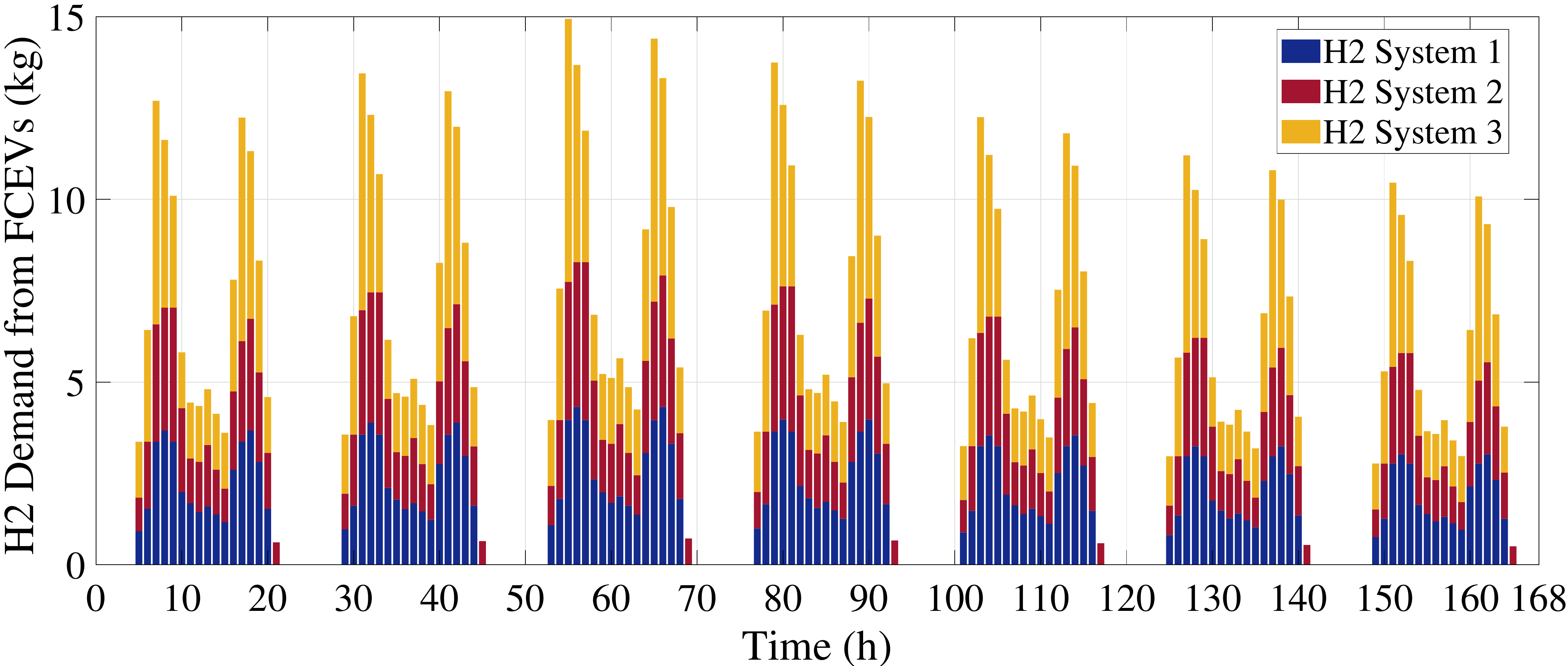}
	\caption{H2 demand from FCEVs}
    \label{HDE}
\end{figure}
\begin{figure}
\centering
\footnotesize
\captionsetup{justification=raggedright,singlelinecheck=false,font={footnotesize}}
	\includegraphics[width=3.5in]{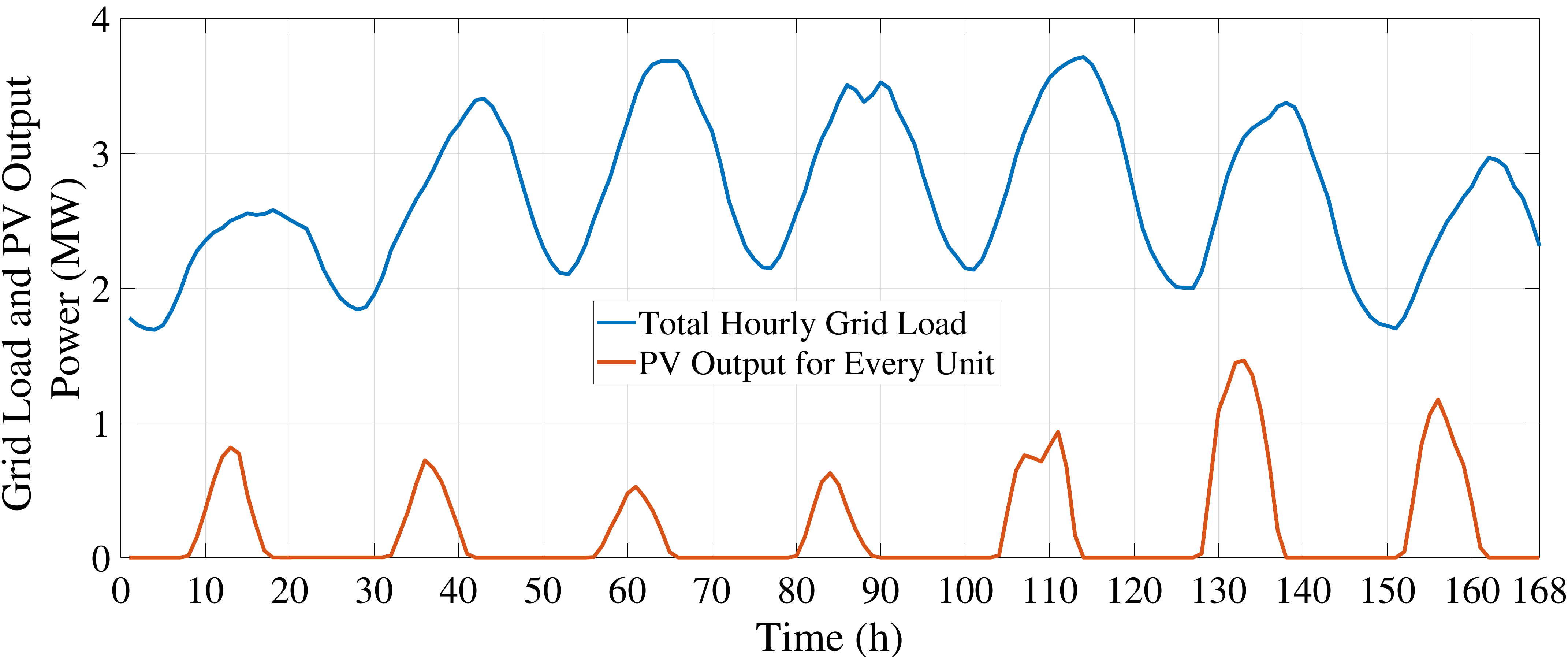}
	\caption{Input data for grid load and output PV power.}
    \label{input}
\end{figure}

\vspace{-0.4cm}
\section{Simulation Results and Analysis}\label{results}
The proposed proactive scheduling of H2 systems is validated on 33-node DN \cite{baran1989network} with an hourly time step for a week. The DN hosts three utility-operated DGs, located at nodes 8, 13, and 30 with capacity limits of 0.8MW, 2.4MW, and 1 MW, respectively. DG 8 and DG 13 are considered as combined-cycle units , and DG 30 is considered as a combustion turbine unit. Additionally, six PV units and three H2 systems are hosted by the DN as depicted in Fig. \ref{system}. The operational costs for generation units are based on \cite{vimmerstedt20182018} for year 2050. H2 demand from FCEVs is calculated based on the method presented in \cite{sun2018optimal}, with the assumption of 20\% penetration level for year 2050. FCEVs are Honda Clarity models \cite{FCEV} assuming that these cars arrive at H2 stations with 45\% fuel in their tank. In other words, FCEVs fill the tanks for the remaining 55\% tank capacity, as depicted in Fig. \ref{HDE}. Considering the extreme event scenario from generation perspective, it is assumed that the tie line between DN and upper grid as well as three DGs are out of service for almost two days (from hour 79 till 128). The optimization time horizon is one week with hourly time steps. The input data for load and every PV unit are shown in Fig. \ref{input}. Moreover, the penetration of PV units is increased by a factor of 2.5 yielding a solar penetration around 40\%. Finally, the parameters used for modeling the H2 systems are presented in table \ref{parameters}. \par 

\begin{table}[]
\centering
\footnotesize
\captionsetup{labelsep=space,font={footnotesize,sc}}
\caption{ \\ Parameters Related To H2 systems }
\label{parameters}
\begin{tabular}{|c|c|c|c|}
\hline
\textbf{Parameter} & \textbf{Value} & \textbf{Parameter} & \textbf{Value} \\ [0.1 cm] \hline

$\eta_{EL}$           & 60\%            & $\eta_{FC}$           & 70\%            \\ [0.05 cm] \hline
$P^{EL,max}$           & 3 (MW)           & $P^{EL,min}$            & 0 (MW)           \\ [0.05050505 cm] \hline
$P^{FC,max}$           & 3 (MW)           & $P^{FC,min}$            & 0 (MW)           \\ [0.050505 cm] \hline
$MOH^{max}$            & 600 (kg)         & $MOH^{min}$            & 60 (kg)          \\ [0.0505 cm] \hline
$\lambda_{Dsp}$               & $0.006 \% \times MOH_{i,t}$     & Specific Energy    & 56.4 (kwh/kg)   \\ [0.05 cm] \hline
\end{tabular}
\end{table}
\vspace{-0.1cm}
\subsection{Simulation Results for Proactive Scheduling of H2 Systems considering FCEVs Demand} 
\begin{figure}
\centering
\footnotesize
\captionsetup{justification=raggedright,singlelinecheck=false,font={footnotesize}}
	\includegraphics[width=3.5in]{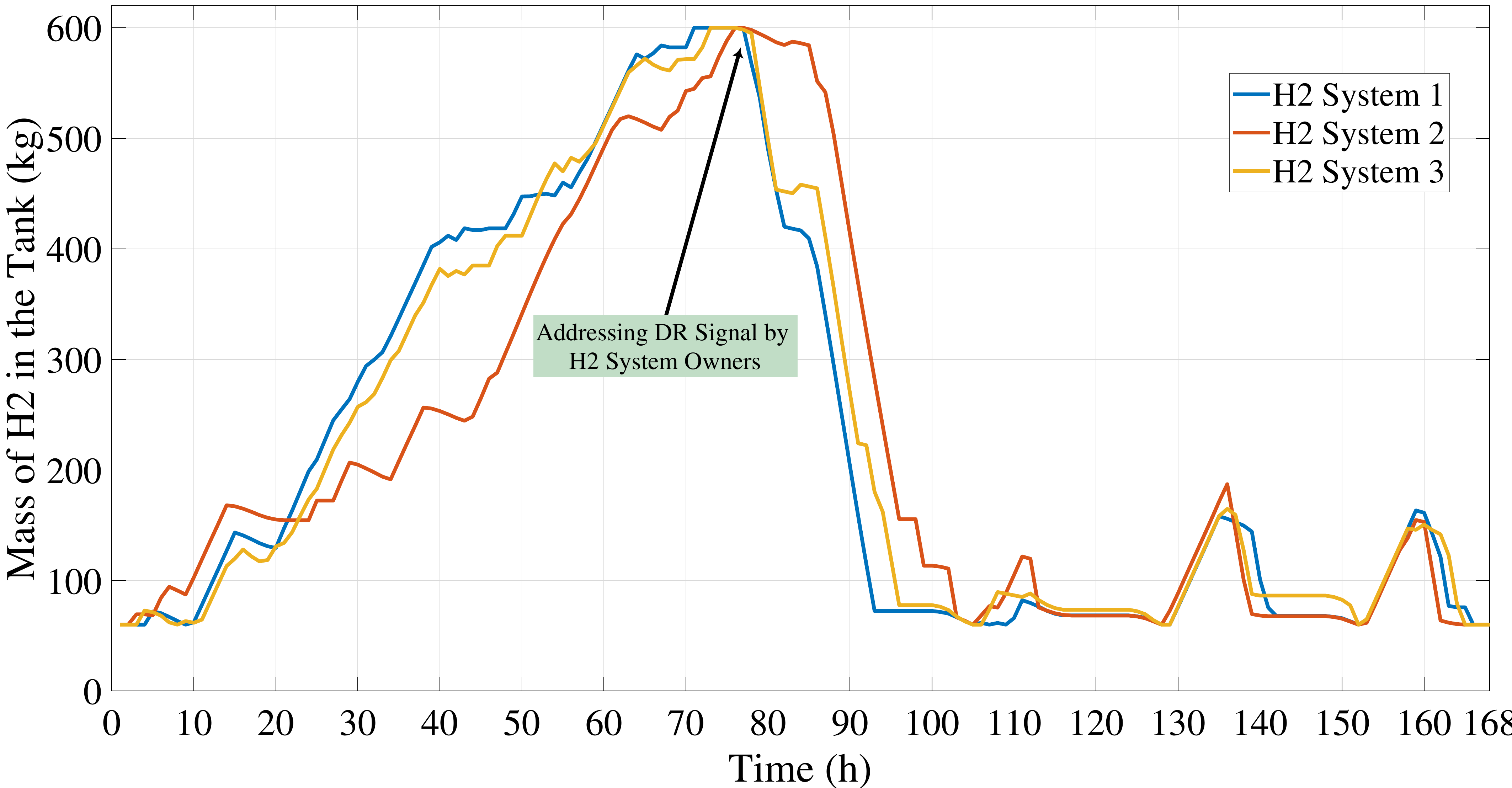}
	\caption{Mass of H2 in the storage tank for H2 systems.}
    \label{LOH}
\end{figure}

\begin{figure}
\centering
\footnotesize
\captionsetup{justification=raggedright,singlelinecheck=false,font={footnotesize}}
	\includegraphics[width=3.5in]{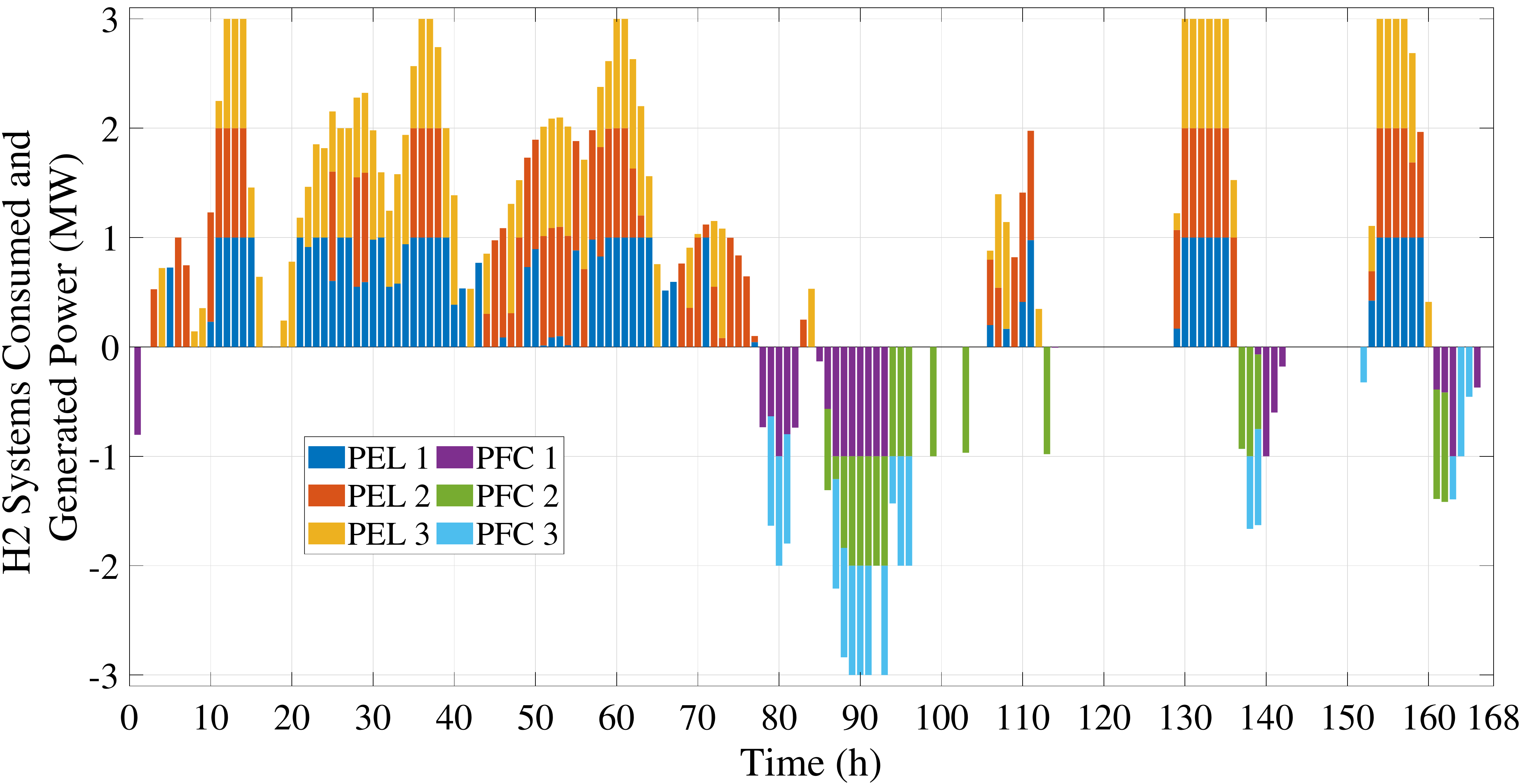}
	\caption{H2 systems consumed and generated power.}
    \label{ELXFC}
\end{figure}

\begin{figure}
\centering
\footnotesize
\captionsetup{justification=raggedright,singlelinecheck=false,font={footnotesize}}
	\includegraphics[width=3.5in]{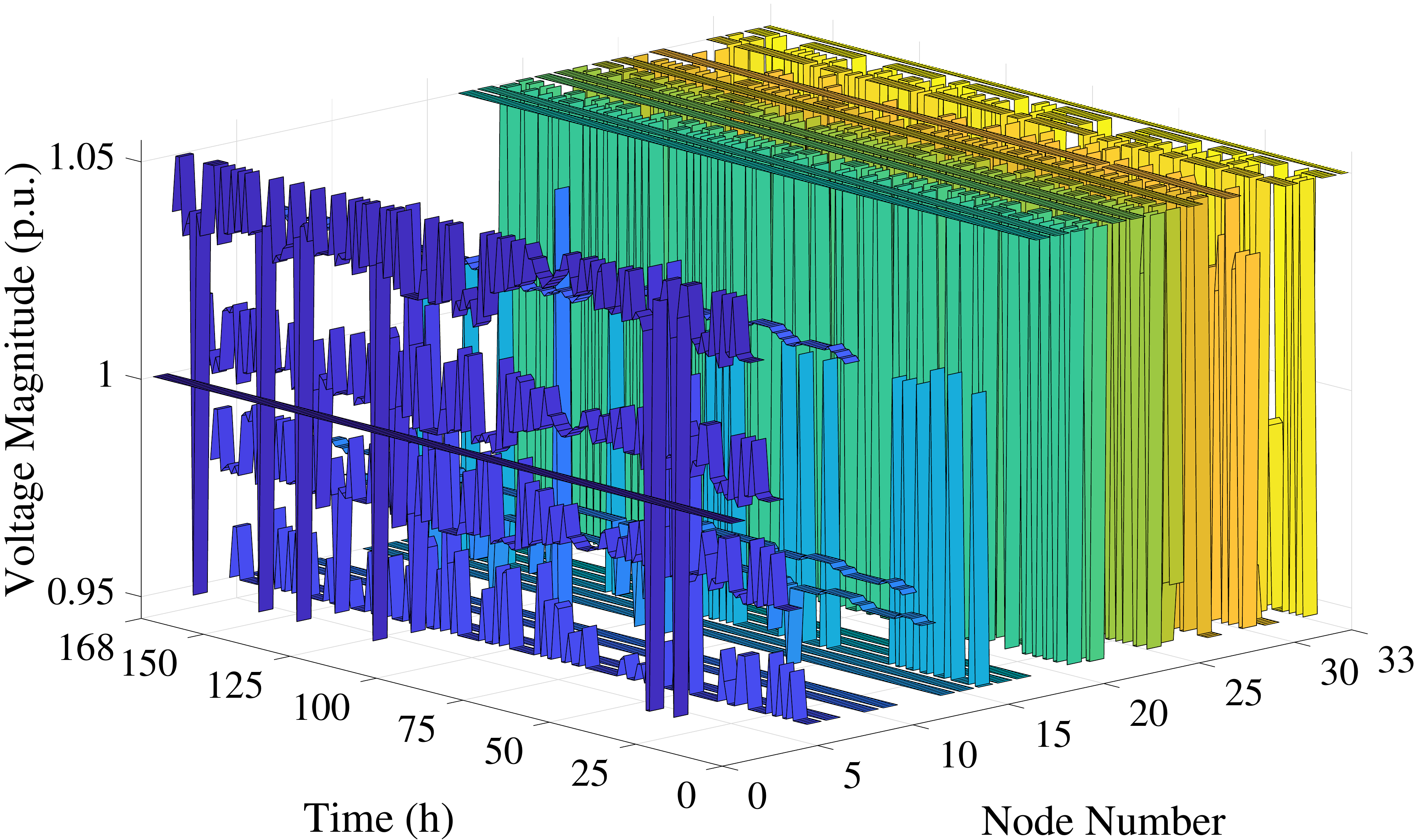}
	\caption{Voltage values during the week.}
    \label{voltage3D}
\end{figure}

The results for proactive management ($t < t_{event}$) and survivability ($t \ge t_{event}$) are shown in Fig. \ref{LOH} to Fig. \ref{voltage}. Fig. \ref{LOH} shows the mass of H2 in the tank during the pre- and post-event times. It shows that based on the signal from DSO, all H2 systems must fill their storage tanks before the event time. As a result, at hour 79, all H2 systems' storage tanks are full. In addition, Fig. \ref{ELXFC} shows the Electrolyzers consumed power prior to the event time to reach the maximum capacity of storage tank. It is also shown that most of the system load is supplied by FC units during hours 79 to 100. Another critical indicator of reliable DN scheduling is voltage values, which the voltage magnitudes and average voltage values are shown in Fig. \ref{voltage3D} and Fig. \ref{voltage}, respectively.\par

\begin{figure}
\centering
\footnotesize
\captionsetup{justification=raggedright,singlelinecheck=false,font={footnotesize}}
	\includegraphics[width=3.5in,height=1.7in]{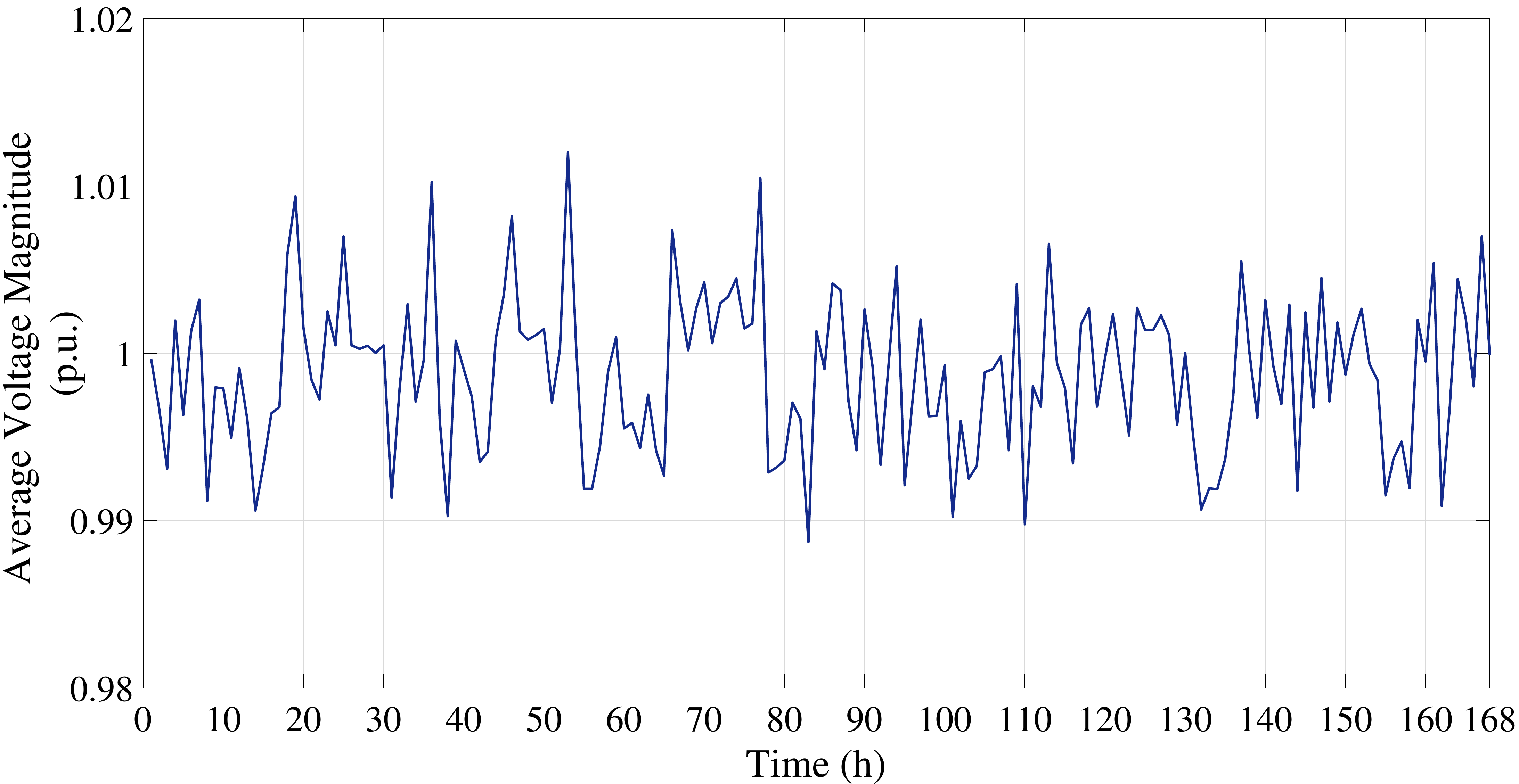}
	\caption{Average voltage values during the week.}
    \label{voltage}
\end{figure}

\subsection{Resilience Analysis for Different Case Studies} In order to provide a fair comparison between FC units and battery storage systems, H2 demand is considered zero kg. Additionally, the power rating of battery power rating is assumed to be equal to that of electrolyzer and FC units. To analyze the resilience of DN with proactive scheduling, six scenarios are considered in handling the out-of-service tie line between DN and upper grid, as well as the three DGs being out of service for almost two days (from hour 79 till 128) The first scenario is the base case without proactive scheduling; next four scenarios are the base case with battery of 2, 4, 6, 8 hours duration, respectively; and the last scenario is the base case with H2 systems including electrolyzer, tank, and stationary FC units. Table \ref{resilience index} shows the results of the unserved energy and the resilience index. It should be noted that resilience index (RI) used in this paper is defined as follows:
\begin{equation}\label{ResilienceIndex}
RI\;(\%)= \Bigg(\frac{Total\;Load - Curtailed\;Load}{Total\;Load}\Bigg) \times 100
\end{equation}
Using the RI, it can be easily seen that maximum unserved energy happened in case 1 (worst scenario) and minimum unserved energy happened when H2 systems are used with stationary FC units as backup generation units. Additionally, higher RI values show that H2 systems can perform better compared to batteries with 2-8 hours duration time, which demonstrates the value for large deployment of these systems in the near future.
\begin{table}
\centering
\footnotesize
\captionsetup{labelsep=space,font={footnotesize,sc}}
\caption{ \\ Resilience Analysis for Different Case Studies}
\label{resilience index}
\begin{tabular}{|c|c|c|c|c|c|c|}
\hline
\textbf{Case Study} &
  \textbf{ 1} &
  \textbf{ 2} &
  \textbf{ 3} &
  \textbf{ 4} &
  \textbf{ 5} &
  \textbf{ 6} \\ [0.1 cm] \hline
\textbf{\begin{tabular}[c]{@{}c@{}}Energy Not Supplied \\ (MWh)\end{tabular}} &
  101.9 &
  91.8 &
  85.3 &
  79.8 &
  75.6 &
  52.0 \\ \hline
\textbf{\begin{tabular}[c]{@{}c@{}}Resilience Index \\ (\%)\end{tabular}} &
  77.4 &
  79.6 &
  81 &
  82.3 &
  83.2 &
  89.2 \\ \hline
\end{tabular}
\end{table}
\vspace{-0.12cm}
\section{Conclusion and Future Work}\label{conclusion}
In this paper, a proactive scheduling of H2 systems in an active distribution network for resilience enhancement is proposed with the goal of minimizing operational cost and load curtailment. Unlike battery storage systems which can only store energy for 4-8 hours, H2 systems including electrolyzers, storage tanks, and FC units can store H2 energy for long duration. In the case of multiple DG and line outages for more than 10 hours, H2 system can significantly enhance the resilience of system by supplying the loads based on their storage tank and FC units. The future work will be the extension of proposed framework by 1) addressing DERs' uncertainty with robust techniques in unbalanced DNs; and 2) analyzing the impact of long-duration H2 storage in resilience enhancement of integrated transmission and distribution systems. 

\section{Acknowledgments}
This work is supported by U.S. Department of Energy's award under grant DE-EE0008851. \par

\bibliographystyle{IEEEtran}
\bibliography{mybib}

\end{document}